\documentclass[fleqn,usenatbib]{mnras}

\usepackage{newtxtext,newtxmath}
\usepackage[normalem]{ulem}

\usepackage[T1]{fontenc}
\usepackage{ae,aecompl}

\usepackage{graphicx}	
\usepackage{amsmath}	
\usepackage{bm}         
\usepackage[capitalize]{cleveref}
\usepackage{algpseudocode} 

\renewcommand{\d}{{\rm d}}

\newcommand{\lM}{\mathcal{M}}
\newcommand{\lD}{\mathcal{D}}
\newcommand{\lZ}{\mathcal{Z}}

\newcommand{\lP}{\mathcal{P}}

\newcommand{\vk}{\bm{k}}
\newcommand{\vp}{\bm{p}}
\newcommand{\vq}{\bm{q}}
\newcommand{\vr}{\bm{r}}

\renewcommand{\vv}{\bm{v}}

\newcommand{\hr}{\bm{\hat{r}}}

\newcommand{\supX}[2]{{#1}^\text{\tiny #2}}

\newcommand{\subX}[2]{{#1}_\text{\tiny #2}}
\newcommand{\snl}{\subX{\sigma}{NL}}
\newcommand{\vnl}{\supX{\vv}{NL}}

\newcommand{\zcos}{\bar{z}}

\newcommand{\LCDM}{$\Lambda$CDM }

\newcommand{\Mpc}{{\rm Mpc}}

\newcommand{\Mpch}{h^{-1}\Mpc}

\newcommand{\kms}{{\rm km/s}}

\newcommand{\mean}[1]{\big< #1 \big>}

\newcommand{\Hamlet}{\textsc{Hamlet}}

\newcommand{\iu}{{i\mkern1mu}}

\title[HMC reconstruction from peculiar velocities]{Hamiltonian Monte Carlo reconstruction from peculiar velocities}
\author[]{
	Aur\'elien Valade,$^{1, 2}$
    Yehuda Hoffman,$^{3}$
    Noam I Libeskind,$^{1, 2}$ 
	and 	Romain Graziani,$^{4}$\\
	$^{1}$Leibniz-Institut für Astrophysik Potsdam (AIP), An der Sternwarte 16, 14482 Potsdam, Germany \\
	$^{2}$Univ Lyon, Univ Claude Bernard Lyon 1, CNRS, IP2I Lyon / IN2P3, IMR 5822, F-69622, Villeurbanne, France \\
	$^{3}$Racah Institute of Physics, Hebrew University, Jerusalem 91904, Israel \\
	$^{4}$OSE Engineering, 1 route de Versailles, 78470 Saint-Remy-Les-Chevreuses, France
}

\date{Accepted XXX. Received YYY; in original form ZZZ}

\pubyear{2021}

\begin{document}

\label{firstpage}
\pagerange{\pageref{firstpage}--\pageref{lastpage}}
\maketitle

\begin{abstract}

The problem of the reconstruction of the large scale density and velocity fields from peculiar velocities surveys is addressed here within a Bayesian framework by means of Hamiltonian Monte Carlo (HMC) sampling. The HAmiltonian Monte carlo reconstruction of the Local EnvironmenT (\Hamlet) algorithm is designed to reconstruct the linear large scale  density and velocity fields in conjunction with the undoing of lognormal bias in the derived distances and velocities of peculiar velocities surveys such as the Cosmicflows data. The \Hamlet\ code has been tested against Cosmicflows  mock catalogs consisting of up to $3\times10^4$ data points with mock errors akin to those of the Cosmicflows-3 data, within the framework of the \LCDM\ standard model of cosmology.

The \Hamlet\ code outperforms previous applications of Gibbs sampling MCMC reconstruction from the Cosmicflows-3 data by two to four orders of magnitude in CPU time. The gain in performance is due to the inherent higher efficiency of the HMC algorithm and due to parallel computing on GPUs rather than CPUs. This gain will enable an increase in the  reconstruction of the large scale structure from the upcoming Cosmicfows-4 data and the setting of constrained initial conditions for cosmological high resolution simulations.

 \end{abstract}




\section{Introduction}
\label{sec:intro}
The large scale structure (LSS) of the Universe, as manifested by the large scale density and velocity fields of galaxies and of the underlying dark matter (DM), reflects the primordial state of the Universe and serves as a probe of the cosmological parameters that define the standard model of cosmology   \citep{Peebles1980,2008cosm.book.....W}. 
In the standard cosmological model the density and velocity fields are strongly related by the continuity equation and in principle  knowledge of one determines the other, in particular in the linear regime. Galaxies are the most common tracers of the  LSS   and   are used to uncover, or reconstruct, the  large scale density and velocity fields. 
Yet, galaxies are biased tracers of the underlying density field \citep{Dekel1986}. Their radial peculiar velocities (namely the residual velocity when the Hubble expected expansion is subtracted from the total velocity) are induced by the density deviations from a homogeneous and isotropic universe. It follows that the velocity field derived from the velocities of galaxies can be used to  uncover the underlying matter, and in particular the DM, distribution and thereby shed light on the so-called galaxy bias. 

The recognition that at low redshifts, $z \lesssim 0.1$, peculiar velocities are practically the only unbiased tracer of the full density  field motivated large scale surveys of galaxy peculiar velocities starting with \citet{1982ApJ...258...64A}, followed by the Seven Samurai discovery of the Great Attractor \citep{1988ApJ...326...19L} and more recently the three data releases of the  Cosmicflows project \citep[CF1, CF2 and CF3, respectively]{2008ApJ...676..184T,2013AJ....146...86T,2016AJ....152...50T}. The availability of peculiar velocity surveys prompted the efforts to reconstruct the continuous density and three dimensional (3D) velocity fields spanned on a regular grid \citep[e.g.][]{Bertschinger1989,1999ApJ...520..413Z,2018NatAs...2..680H,2016MNRAS.457..172L}. The reconstruction of the LSS from peculiar velocity surveys is challenging - the data is sparse, non-uniform, anisotropic, extremely noisy and biased. Yet, the highly correlated nature of the  velocity field, predicted  by the standard model, acts to counter balance these hurdles and makes the reconstruction of the LSS possible.

The radial peculiar velocity of a  galaxy is not a directly observed quantity but is  an inferred one. (Throughout the paper the term `velocity' stands for the radial peculiar velocity.)  In fact surveys of galaxy velocities are actually surveys of galaxy distances and redshifts, from which  velocities are inferred. Yet,  galaxy distances are inferred from the observed distance moduli. The normally distributed observational uncertainties of the  distance moduli are transformed into lognormal errors  and thereby to a bias in the estimated  distances and velocities of galaxies. This lognormal bias is traditionally associated with the Malmquist bias that severely affects the analysis of galaxy  velocities surveys \citep{1995PhR...261..271S,2021MNRAS.505.3380H}. The relative distance errors in the Cosmicflows database are predominantly  of the order of $\sim20\%$, which translate to velocity errors of that order of the Hubble velocity. For a galaxy at a distance of $100\,\Mpch$ say (where $h$ is Hubble's constant expressed expressed in units of $100\,\kms/\Mpc$) the typical uncertainty on the velocity is $2000\,\kms$. In the \LCDM\ standard model of cosmology with the Planck parameters \citep{Ade2016} the predicted scatter of field galaxies is $\sim350\,\kms$.  This sets the stage for the enormous challenge one faces in attempting to recover the full density and velocity fields from velocities surveys - from sparse, nonuniform, anisotropic and  biased data with a typical noise over signal ratio as small as $\sim15\%$.  Another obstacle to overcome is that only the radial component of the velocity is estimated yet a reconstruction of the 3D velocity vector field is needed.

It is clear that the data itself cannot determine the density and velocity fields even within the data zone, namely the regions of space that are sampled by the data, let alone outside it. Something else, beyond the data, is needed in order to infer from the radial  component of the velocity their three Cartesian components and the associated density field. This was done first by POTENT algorithm \citep{Bertschinger1989}, which assumed the linear theory of gravitational instability in an expanding universe. Yet no assumption was made by POTENT on the statistical nor on the random nature of the underlying density and velocity field. 

One of the basic tenets of the \LCDM\ model is that the  primordial perturbation field constitutes a random Gaussian field, and as such its properties are determined by its power spectrum. Furthermore, in its linear regime the density and velocity fields are linearly related and the knowledge of one determines the other. This has led to a Bayesian approach to the reconstruction of the LSS that relies explicitly on the Gaussianity of the density and/or velocity fields by meas of the Wiener filter (WF) and constrained realizations (CRs) of Gaussian random fields \citep{1991ApJ...380L...5H,1995ApJ...449..446Z, 1999ApJ...520..413Z,2016MNRAS.455.2078S,2021MNRAS.505.3380H}.
A later development has been made by employing by means Markov Chains Monte Carlo (MCMC) sampling  \citep{2016MNRAS.457..172L,Graziani2019}. 

In both approaches, the WF/CRs and the MCMC, the estimation of the desired density and velocity fields proceeds by the estimation of their posterior probability distribution functions (PDFs) given the data and the assumed \LCDM\ model. Yet, the two methods treat uncertainties that lie outside the formalism of Gaussian random fields in radically different ways. This is most strongly manifested by the way the lognormal bias is treated. In the WF/CRs approach the correction of the longnormal bias is done outside and independently of the WF/CRs analysis. Namely a procedure is applied to correct the bias, yielding bias corrected estimated distances and velocities of the data points, and then the WF/CRs machinery is applied to the correct data (\citealt{2015MNRAS.450.2644S} and independently in \citealt{2021MNRAS.505.3380H}). Such an approach enables the linear reconstruction   from the biased-corrected data by means of an exact analytical solutions. The MCMC adopts a holistic nature to the problem and combines all the uncertainties of the data into one  PDF. In particular, statistical assumptions are  made on the distribution of the distances of the data points and on the errors of the observed distance moduli. Such an approach defies an exact analytical solutions and the  estimation is done by means of MCMC sampling. Gibbs sampling is used by \citet{2016MNRAS.457..172L} and by \citet{Graziani2019}.

The aim of the present paper is to present a new numerical approach to the MCMC algorithm of \citet{2016MNRAS.457..172L} and by \citet{Graziani2019}. The Gibbs sampling algorithm suffers from a slow convergence and is very CPU inefficient in comparison with the WF/CRs formalism. A very considerable improvement is presented here by a numerical implementation of the Hamiltonian Monte Carlo (HMC) sampling technique.
The HMC sampling technique was applied before to reconstruct the LSS from a galaxy redshifts survey by \citet{2019A&A...625A..64J} and it is applied here for the first time to a galaxy velocities survey. Both the Gibbs sampling based MCMC reconstruction implementations and the present HMC one  are constructed within the same theoretical framework ans to the extent that they are applied to the same data aiming at the same resolution they should yield very similar results. Our motivation here is  to considerably improve the numerical efficiency of MCMC algorithm, aiming in particular to achieve the numerical resolution needed for setting up constrained initial conditions for numerical simulations of the local universe \citep[e.g.][]{2014NewAR..58....1Y,2016MNRAS.455.2078S,Libeskind2020}. The HAmiltonian Monte carlo reconstruction of the Local EnvironmenT (\Hamlet\ ) code is presented here and is tested against a mock catalog. The comparison of the application of the \Hamlet\ code and a WF/CRs reconstruction is presented in  Valade, Hoffman and Libeskind (in prep).

The paper starts with a detailed description of the HMC methods (\cref{sec:HMC}). 
A description of the application of the \Hamlet\ code to a mock velocity survey and its analysis follows (\cref{sec:mock}). 
A summary and discussion section concludes the paper (\cref{sec:summary}).

\section{Bayesian inference of the large scale structure}
\label{sec:Bayes}

The Bayesian probabilistic approach is adopted here, according to which the PDF represents one's confidence in the certainty of the knowledge of the values of some  parameters/variables. These  can be either observables that can be measured or theoretical parameters whose values are to be inferred. The aim of the Bayesian analysis, in the present context, is the construction of the posterior PDF of the distances of the data points, the set of the Fourier modes that define the density and velocity fields, and $\snl$, given Cosmicflows-like data and under the assumption of the \LCDM\ model.
More specifically, the posterior PDF is calculated within the linear approximation of the \LCDM\ model, with the Planck parameters \citep{Ade2016}. A mild non-linear correction is added to the linearly calculated velocity field whose amplitude is controlled by $\snl$ (to be defined later), whose magnitude is to be estimated as well.

 The Bayesian posterior PDF is numerically evaluated by means of the MCMC sampling. Given the posterior PDF the mean and variance of the desired density and velocity fields are readily calculated, much in the same way as in the WF/CRs formalism \citep{1995ApJ...449..446Z}. The main difference between the WF/CRs and the MCMC cases is that in the former  the posterior PDF is assumed to be know analytically and in the later it is evaluated numerically. 

\subsection{Bayesian posterior PDF and likelihood function}
\label{sec:posterior-like}

Bayes' theorem states that the  posterior PDF, of the model given the observed data, is the product of the conditional probability of the data given the model, hence the likelihood function, times the prior probability of the model, normalized by the evidence. In the language of Bayes' theorem the distribution of the true distances of the data points, the ensemble of the Fourier modes and  $\snl$ consist the multi-parameter model whose parameters are to be estimated given the data. The \LCDM\ cosmological model provides the framework within which that multi-parameter model is constructed. Neglecting here the evidence the posterior PDF is:
\begin{equation}
    \label{eq:covfefe:bay}
    P\big(\Delta_k, \lD, \snl | \lM, \lZ\big) \propto
    L\big(\lM, \lZ | \Delta, \lD, \snl\big) 
    P\big(\Delta, \lD, \snl \big)
\end{equation}
Here $\lD = \{d_i\}$ (proper distances of the data points, $i=1, ...\ n$ where $n$ is the number of data points), $\Delta_k = \{\delta_{\vk} \} $  (the ensemble of the Fourier modes, where $\delta_{\vk}=FT\left( \delta({\vr} )\right)$ is the Fourier transform of the fractional over-density field $\delta(\vr)$) and $\snl$ are the output variables/parameters  to be estimated. $\lZ = \{ z_i\}$ (observed redshifts of the data points) and $\lM = \{\mu_{i}\}$ (observed distance moduli) are the input observed data. $L\big(\lM, \lZ | \Delta, \lD, \snl\big)$ is the likelihood function and $P\big(\Delta, \lD, \snl \big)$ is the prior. 

The errors on the observed data points  are assumed to be independent, and so are the errors on the redshift and distance modulus  of a given galaxy. All observational errors are assumed to be normally distributed. The likelihood function  is thus the product of $n$ likelihood functions, one for
each of the $i$ constraints used. Furthermore, given that the redshifts and distance moduli are independent, the $i$-th likelihood function is a product of two independent likelihood functions denoted here by   $L_i^{v_r}$, associated eventually with the  velocity of the data point (\cref{eq:vrobs} below)  and  $L_i^{\mu}$. The  likelihood function is:
\begin{equation}
L\big(\lM, \lZ | \Delta, \lD, \snl\big)
                  = \prod^n_i L_i^\mu( \mu_i | \Delta, \lD, \snl ) L_i^{v_r}( z_i | \Delta, \lD, \snl)
\label{eq:likelihood}
\end{equation}

\subsection{The Likelihood function: distances}
\label{sec:dist}

Given a Gaussian error $\sigma_{\mu,i}$ on the measurement, $L_i^\mu$ is written:
\begin{equation}
    L_i^\mu(\mu_i | d_i) = \frac{1}{\sqrt{2 \pi \sigma_{\mu, i}^2}}
    \exp \left( \frac{-\big(\mu_i - \mu(d_i)\big)^2}{2 \sigma_{\mu, i}^2}  \right) 
\label{eqn:lmu}
\end{equation}
where
\begin{equation}
    \mu(d) = 5 \log_{10}\left(d_L(d) \over 10 {\rm pc}\right),   
\end{equation}
where $d_L(d)$ is the luminosity distance associated with the proper distance $d$.
\begin{equation}
   d_L(d)=d  \big(1 + \zcos(d)\big)
\end{equation}
and to 2nd order the cosmological redshift ($\zcos$) corresponding the proper distance $d$ is given by:
\begin{equation}
    \zcos(d) = \frac{2}{3 \Omega_0} \left( 1 - \sqrt{1 - \frac{3 \Omega_0 H_0}{c} d} \right)
\end{equation}
Here $\Omega_{0}$ is the matter density parameter, $H_{0}$ is Hubble's constant evaluated at the present epoch and $c$
is the speed of light. 

\subsection{The Likelihood function: velocities}
\label{sec:vel}

The velocity of the $i$-th  data point is related to its observed redshift  via
\begin{equation}
v_{r,i}^{\rm obs}  = c\frac{z_i - \zcos(d_i)}{1 + \zcos(d_i)}.
\label{eq:vrobs}
\end{equation}
Given the ensemble of Fourier modes, $\Delta_k$,  the assumed cosmological  3D velocity field is given by the following inverse Fourier transform,
\begin{equation}
    \vv(\vr | \Delta_k)=FT^{-1}\left(-\iu H_0 f(\Omega_0){\vk\over k^2}\delta_{\vk} \right),
    \label{eq:vel-field}
\end{equation}
where $f(\Omega_0)$ is the linear growth factor. One should note here that $\vv(\vr | \Delta_k)$ is the velocity field predicted by the linear theory from a given over-density field $\delta(\vr)$.

Our aim here is to reconstruct the LSS from the grouped version of a Cosmicflows-like catalog of galaxy velocities. A grouped catalog means here a  catalog in which all galaxies belonging to a group or cluster of galaxies are collapsed onto one data points. The grouping acts as a smoothing process of the internal virial velocities and thereby serve as a filter of non-linear velocities. The  following crude approximation is introduced here to account  for the residual  non-linear component of the observed data. The full velocity field is assumed to include a non-linear component:
\begin{equation}
    \vv^{\rm full}(\vr) = \vv(\vr | \Delta_k) + \vnl(\vr)
    \label{eq:v-full}
\end{equation}
The non-linear component, $\vv^{nl}(\vr)$, is assumed to constitute a white noise with a variance given by:
\begin{equation}
    \snl^2=\left< \vnl(\vr)^2\right>.
\end{equation}

The  likelihood function for the  observed redshift is readily written here 
in terms of the observed velocity:
\begin{equation}
    L_i^{v_r}( z_i |  \Delta_k, \lD, \snl ) = \frac{1}{\sqrt{2 \pi} \kappa_i} 
    \exp \left( \frac{-\left(v_{r,i}^{\rm obs} - \vv(d_i\hr_i | \Delta_k)\cdot\hr_i\right)^2}{2 \kappa_i^2}  \right) \\
\label{eqn:lvr}
\end{equation}
where $\hr_i$ is the unit vector in the direction of the $i$-th data point and
\begin{equation}
    \kappa_i^2 = \snl^2 + \frac{\sigma_{cz, i}^2}{(1 + \zcos(d_i))^2}. 
    \label{eq:kappa}
\end{equation}

\subsection{Priors}

The elements of the model under consideration are the Fourier modes, the distribution of the distances of the data points and $\snl$. Hence the
joint prior of the model is written as:
\begin{equation}
    P(\Delta_k, \lD, \snl) = P(\Delta_k) P(\lD) P(\snl).
\end{equation}

The Fourier modes are evaluated on a discrete grid which is written here symbolically as $\left\{\vk_j\right\}$, where $j=1, ... , m$ where $m$ is the number of the Fourier modes. 
The prior of the ensemble of the Fourier modes is:
\begin{equation}
    P(\Delta_k) = \prod_{j=1}^{m} \frac{1}{2\lP(k_j)} \exp \left( \frac{- |\delta_{\vk_j}|^2}{\lP(k_j)} \right)
    \label{eq:prior-k}
\end{equation}
Here  $\lP(k)$ is the \LCDM\  power spectrum at wave number $k$. Writing the prior on the
distances is  complicated and somewhat ad hock  \citep[see extensive description in][]{Graziani2019}. We opt for 
a simple  description based on the fact that the redshift distance, $d_z$  is a good proxy to the actual distance for all but the very nearby data points:
 \begin{equation}
    d_z = \frac{c}{H_0} \int_0^z\frac{\d \tilde{z}}{\sqrt{\Omega_m (1 + \tilde{z})^3 + (1 - \Omega_m)}}.
    \label{e:dz}
\end{equation}
A histogram of the distribution of the  redshift distances,   $N_{d_z}^{\rm obs}$ is constructed from the   distribution of the observed redshifts and is taken to be a good proxy to the histogram of the distribution of the true distances. Hence, the prior of the true distance is approximated by: 
\begin{equation}
    P(\lD) = \prod_{i=0}^{n} N_{d_z}^{\rm obs}(d_i).
    \label{eq:D}
\end{equation}

\section{Hamiltonian Monte Carlo sampling}
\label{sec:HMC}

Given the posterior PDF (\cref{eq:covfefe:bay}) the inference of the LSS from the data and the assumed prior is well posed and in principle can be readily calculated. Yet, the treatment of the distances  of the data points as part of the parameters of the model to be inferred makes the problem of inference intractable analytically. This leaves numerical sampling of the posterior PDF as the only viable way for calculating the mean and variance, say, of the density and velocity fields given the data and the prior. The very large numbers of the parameters of the model, the so-called the curse of dimensionality, calls for sophisticated sampling methods, where the sampling is done by random Markov chains generated by random Monte Carlo realizations. The construction of Markov chains often follows the the   Metropolis-Hastings or   the Gibbs  algorithms. Yet, the numerical efficiency and the slow convergence of these algorithms limit their applicability to very high dimensional cases.

The   HMC   is an   algorithm that enables   an efficient sampling of complex  posterior PDFs of high dimensionality models and large input databases. There is ample evidence in the literature  that the exploration provided by HMC methods is order of magnitudes more efficient than  the  Metropolis-Hastings or even the Gibbs sampling. The reader is referred to the excellent  review of the HMC method and its comparison with other MCMC sampling methods of \citet{2011hmcm.book..113N} and in particular to \citet{2010MNRAS.407...29J} and \citet[and references therein]{2019A&A...625A..64J}   for the applications of HMC to the reconstruction of the LSS from redshift surveys.

\subsection{Hamiltonian trajectories in phase space}
\label{sec:trajectory}

A detailed presentation of the MCMC sampling in general and the HMC sampling in particular  of the posterior PDF is beyond the scope of the paper. Only the key general elements of the HMC are given here. The parameters of the model are denoted here by:
\begin{equation}
    \vq  = \{q_a\} 
            = \big(\delta_{k,1}^R,\, \hdots ,\,\delta_{k,m}^R,\,
                \delta_{k,1}^I,\, \hdots ,\,\delta_{k,m}^I,\,
                d_0,\, \hdots,\, d_n,\, \snl\big)  
                \label{eq:q}
\end{equation}
Here, the real and imaginary components of the complex $\delta_k=\delta^R_k+i\delta^I_k$ are denoted as separate parameters. For the sake of the clarity of the presentation we define here the posterior function as $\Pi(\vq)=P\left(\Delta_k, \lD, \snl | \lM, \lZ\right)$. 

A Hamiltonian system is defined by means of associating $\Pi(\vq)$ with a potential of classical particles, $\Psi(\vq)$,
\begin{equation}
    \Psi(\vq) = -\ln\Pi(\vq).
    \label{eq:Psi}
\end{equation}
The parameters $\{q_a\}$ are assumed to consist a set of canonical coordinates. These are supplemented by auxiliary quantities, referred to as their associated momenta, $\vp  = \{p_a\}$  and a ’mass matrix’ $M$. The dynamics of this Hamiltonian system is governed by the Hamiltonian,
\begin{equation}
    H\left(\vp, \vq\right) = \frac{1}{2} \vp^T M^{-1} \vp + \Psi(\vq)
    \label{eq:Hamiltonian}
\end{equation}
The equations of motions of the $\vq$ and $\vp$ are given by Hamilton equations:
\begin{equation}
    \begin{split}
         \frac{\d p_a}{\d t} &= - \frac{\partial H}{\partial q_a} = -\frac{\partial \Psi(\vq)}{\partial q_a}, \\
    \frac{\d q_a}{\d t} &= \frac{\partial H}{\partial p_a}
    \end{split}
    \label{eq:hamilton}
\end{equation}

Considering these Hamiltonian system as a many body system, probability distribution function  of $\vq$ and $\vp$, $\Pi(\vq,\vp)$ is related the the Hamiltonian via:
\begin{equation}
    \Pi(\vq,\vp) \propto \exp{\left(-H(\vq,\vp)\right)}=\Pi(\vq)\exp{\left(-\frac{1}{2} \vp^T M^{-1} \vp\right)}. 
    \label{eq:Hexp}
\end{equation}
This is a key result. There is no cross-correlation between the distributions of the coordinates and the momenta in the  joint PDF $\Pi(\vq,\vp)$. It follows that the Hamiltonian trajectories  properly sample the desired posterior PDF of the parameters of the model under consideration.

\subsection{Construction of HMC chains}
\label{sec:HMC_chain}

A chain starts with the coordinates and momenta randomly drawn and serve as the initial conditions for the integration of equations of motion (\cref{eq:Hamiltonian}). These are integrated over a   pseudo time $\tau$. The final position of that trajectory in the $\{\vq, \vp\}$ phase space, is the candidate state.  If endpoint failed the standard Metropolis- Hastings acceptance rule, the integration starts all over with the same initial coordinates but with new randomly drawn momenta.
If the endpoint passed  the   acceptance rule, that trajectory becomes a step along the chain. The next step starts with the coordinates ($\vq$) accepting the final coordinates from the last step and the momenta ($\vp$), on the other hand, are again randomly drawn. The final positions of the successive trajectories form the chain.

\subsection{Integrating the HMC trajectories}

If the integration can be done analytically, the Hamiltonian framework insures an acceptance rate of
1, namely all candidate states are accepted, even when candidates  are far to the current position.
However, for most problems, analytical integration is impossible and numerical solvers must be used.
The integrator of choice is the Leapfrog algorithm, which insures ergodicity, namely the
conservation of the Hamiltonian. This insures that the error introduced when computing the
trajectory depends only on the integration step size and not on the number of integration steps. In
other words, the use of the Leapfrog algorithm yields stable trajectories. In practice, this
stability is limited by the numerical precision of the derivatives $\left(\frac{\partial
\ln~\Pi(\vq)}{\partial q_a}\right)$. Thus, the HMC is only applied to models where these derivatives can
be computed analytically, which is the case for this work. 
 
Note that trajectory length is simply the product of the step size and the number of steps used in the
(leapfrog) integration. In this context, the acceptance rate for a given state is only a function of the step size.
Therefore the step size can be tuned in order to obtain a given acceptance rate. Studies in the
literature \citep[e.g.][]{Hoffman2011} advocate that an acceptance rate of $0.65$ (used here)
provides an optimal balance between computational resource usage and exploration. \cite{Hoffman2011}
introduced the ``Dual Averaging'' method that dynamically tunes the step size to reach any given
acceptance rate. The step size set to achieve this acceptance rate depends on the complexity of the
problem: the more complex the problem and the more correlated the variables, the smaller the step
size must be. 

Secondly, since the integration is ergodic, every trajectory ultimately returns to the initial state
(to within integration error). This is problematic since such closed orbits return a final state
identical to the initial state resulting in no additional knowledge of the parameter space (and a
waste of computational resources in the process). Therefore it is absolutely critical that the
integration is halted after a designated number of steps, specifically chosen such that the
candidate state is the furthest away from the initial position. At this maximum, the trajectory
starts turning back towards the initial state. Depending strongly on the initial momentum, this
value is different for each trajectory. Several methods have been developed to automatically tune
this parameter, the most well known being the No U-Turns (NUTS) algorithm proposed in
\cite{Hoffman2011}, whose detail is out of the scope of this paper. NUTS and the ``Dual averaging'' technique can be used together.

\subsection{The Mass matrix}

Consider the classical dynamics described by \cref{eq:hamilton} - a trajectory evolves from a random position in the multi-dimensional phase space towards a local minimum of the potential $\Psi(\vq)$. In the presence of dissipative forces, the trajectory would reach  a local minimum of the potential and stay there. For a Hamiltonian system whose energy is conserved, the trajectory oscillates around the local minimum with an amplitude dictated by the energy of the system and its mass. As the energy of each trajectory is set by the random choice of its initial momentum. The statistics of these initial momenta is encoded in the mass matrix. The selection of a mass matrix influences heavily the efficiency of the exploration and the rapidity of the convergence. Asymptotically however, it does not bias nor modify the result. Even though there is theoretically no optimal choice of mass matrix, using the covariance of the parameters is the canonical approach. This covariance matrix is however a priori not know and has to be estimated.

In the absence of observational data the posterior PDF degenerates into the prior PDF and the model parameters ($\vq$) are all statistically independent and their statistical behaviour is fully understood. The real and imaginary components of the Fourier modes are normally distributed around zero with a variance given by the power spectrum \citep[see][for further discussion]{2010MNRAS.407...29J}. The proper distances are distributed around the corresponding redshift  distances with a variance given by $(\sigma_V/H_0)^2$ (where $\sigma_V$ is the \LCDM\ one dimensional velocity dispersion). The choice of the size of the other parameters (limited to $\snl$ in this work) has to be estimated more freely. 

The  mass matrix for the trivial case of no observational data, namely the posterior equals the prior PDF, is thus written as:

\begin{equation}
    M^{-1}_{ab} = I_{ab} \times
    \begin{cases}
        \lP(k_a) / 2 & \text{ if } a \le m \\
        \lP(k_{a - m}) / 2 & \text{ if } m < a \le 2 m \\
        \sigma_v^2/ H_0^2 & \text{ if } 2 m < a \le 2 m + n \\
        1 & \text{ if } a = 2 m + n + 1
    \end{cases}
    \label{eq:M_ab}
\end{equation}
where $I$ is the identity matrix and $k_{a}$ is the wavenumber of the Fourier   mode $\delta_{k,a}$. The mass matrix of \cref{eq:M_ab} is used here in the general case, where the estimation is done given the data and the prior.

\subsection{Sampling by HMC chains}

An HMC chain is made up of a set of steps, often hundreds of these, which  mark the endpoints of Hamiltonian trajectories. Typically, the HMC algorithm is “ergodic” — it is not  trapped in some sub-volume of the phase space but has a finite probability of visiting any volume element of phase space. Moreover, the property of detailed balance of the HMC chains \citep[see][]{2011hmcm.book..113N} insures the probability of a chain to visit the volume element $\d\vq$ equals $\Pi(\vq)$. Namely, HMC chains sample the parameters space $(\vq)$ with the desired posterior  probability. This leads to the practical use of the HMC sampling to estimate the parameters of the model given the data and the prior. Consider an HMC chain $S$,  $\vq(s) = \left\{\vq_1, \hdots ,\vq_{n_S}\right\}$, and  some functional of the parameters of the model $F(\vq)$, then the conditional mean value of $F(\vq)$ given the data and the prior is given by:
\begin{equation}
	\mean{F(\vq)\ | {\rm\ data,\ prior}} = \int F(\vq) \Pi(\vq) \d\vq \approx \frac{1}{n_S} \sum_{s=1}^{n_S} F\left(\vq_s\right).
    \label{eq:int_mc}
\end{equation}

A further use of HMC chains is the construction of constrained realizations of the parameters $\vq$, given the data and the prior. Namely, the generalization of constrained realizations of Gaussian random fields \citep{1987ApJ...323L.103B,1991ApJ...380L...5H}  to the general case of non-Gaussian posterior PDFs. This is done by selecting states on a given chain that are well separated along the chain, so to insure their statistical Independence.

\section{Technical Implementation}
\label{sec:technical}

An HMC algorithm computationally outperforms the more traditional  Metropolis - Hastings and Gibbs sampling algorithms. This is also the case with the problem, of the reconstruction of the LSS from a CF3-like catalogue, addressed here and by \cite{Graziani2019} which used the Gibbs sampling approach. Quantifying the speed-up of the  \Hamlet\ method  compared to that of \cite{Graziani2019} is complicated because there are differences to both the implementation (compiled on GPUs versus interpreted on a single CPU) and the algorithm (HMC versus Gibbs sampling). Its thus not straight forward to identify exactly which aspect of the \Hamlet\ method is mostly responsible for the increase in efficiency. A quick comparison shows that the \Hamlet\ code outperforms the code of \cite{Graziani2019} by several orders of magnitude (between 3 to 4) in speed while fitting orders of magnitudes more parameters ($\sim 2\times10^6$ versus $\sim 4.5\times10^4$). For example, in the case of reconstructing the LSS  from the $\sim1.5\times10^4$ constraints provided by the grouped CF3 catalogue, the
MCMC method of \cite{Graziani2019} takes more than a month compared with the \Hamlet\ which takes on the order of 10 minutes. The increase in speed is a necessary condition for future applications of the \Hamlet\ code. One such application is the setup of constrained initial conditions for high resolution cosmological simulations \citep[cf.][]{Libeskind2020}, for which the number of the needed Fourier modes is much larger that what is used here. Also, the upcoming 4th  Cosmicflows data release (CF4) is expected to roughly triple the size of the CF3 data. Preliminary analysis suggests that \Hamlet\ will be capable in exploiting the CF4 data. 

A brief review of the  computational implementation of the \Hamlet\ code follows. It takes advantage of a number of highly-abstract layers as implemented by open-source \texttt{Python} libraries \texttt{tensorflow} and \texttt{tensorflow-probabilities}. The \texttt{tensorflow} library provides a framework that enables a python code to transparently scale on multiple CPUs and/or GPUs and to be compiled at run time, while the \texttt{tensorflow-probabilities} provides a plug-and-play implementation of the HMC, 
NUTS and other tools to run and analyse MCMC chain. While the gradient of the the posterior PDF,  can be extremely tedious to write by hand, \texttt{tensorflow} is  capable of transparently computing it, by constructing a complex derivation graph that can be very efficiently evaluated. Only the gradient of the inverse of the Fourier transform had to be constructed.  

\section{Testing Hamlet against a linear mock Cosmicflows-3 survey}
\label{sec:mock}

The \Hamlet\ algorithm is tested against a CF3-like survey drawn from a linear random realization of the density and velocity fields, constructed within the framework of the \LCDM\ model. The selection of the data points and the assignment of the observational errors replicate the selection and the errors of the actual CF3 data. Our aim here is to test the the \Hamlet\  algorithm and its performance in the ideal case where in the limit of perfect data - densely, homogeneously and and isotropically  sampled and with negligible errors - the \Hamlet\ should accurately recover  the input density and velocity fields \citep[cf.][]{Graziani2019}. Testing \Hamlet\ against a mock non-linear  CF3 database is presented in  Valade, Libeskind and Hoffman (to be submitted)

\subsection{Mock Catalogue construction}

A  linear realization of a Gaussian random field, defined by the \LCDM power spectrum and cosmological parameters  \citep{Ade2016} is used here as a base for our mocks. The field is constructed on a $128^3$ Cartesian grid within a box with side length $L=500\Mpch$. Periodic boundary conditions are assumed. A random observer is selected to reside at the center of the computational box and a mock Supergalactic coordinate system is assigned centered on the observer and aligned with the principal directions of the grid. A mock catalog consists of  Supergalactic latitude (SGB)  and  longitude(SGL), distance modulus
($\mu$), its error ($\sigma_{\mu}$), the redshift ($z$) and its error ($\sigma_{cz}$).

\begin{figure}
    \centering
    \includegraphics[width=\columnwidth]{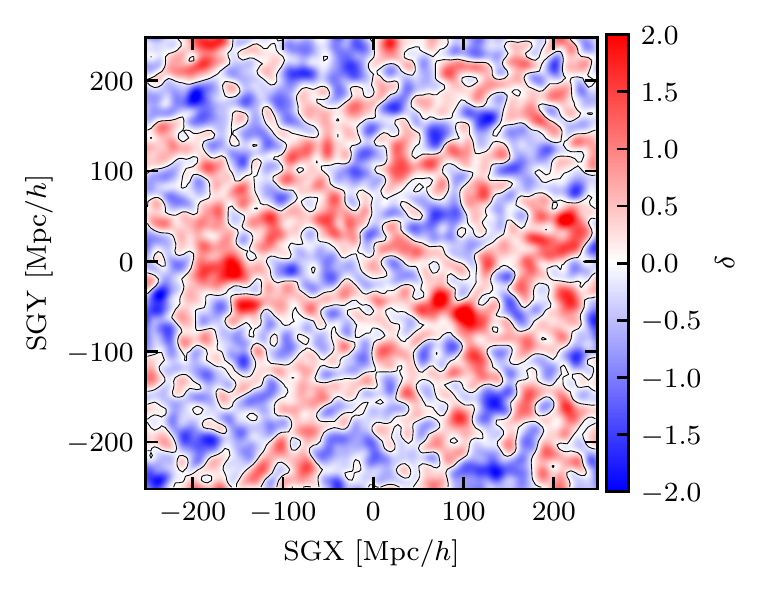}
    \includegraphics[width=\columnwidth]{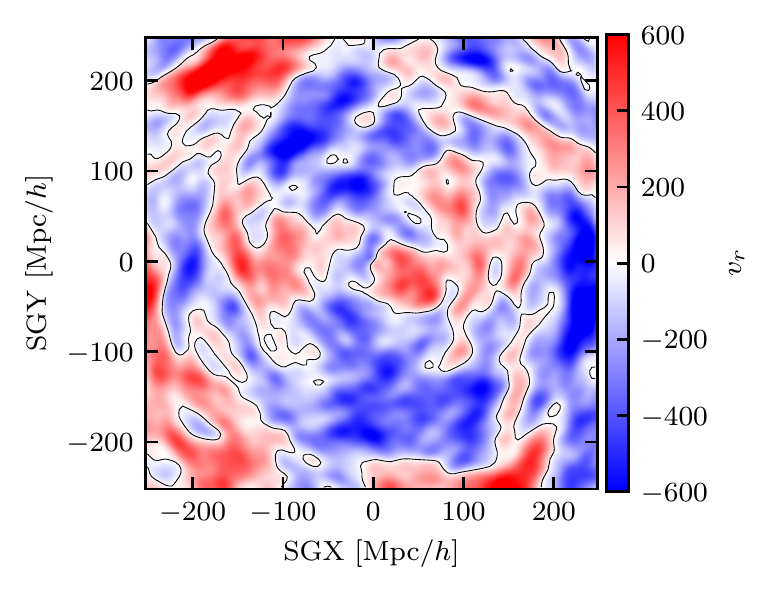}
\caption{   A slice through the over-density ($\delta$,upper panel) and the radial component of the velocity fields  ($v_{\rm r}$, lower panel) of the target field. The mock observer is located at the origin of the coordinate system. The color bars present the color coding of the presented fields. Velocities are in units of $\kms$. The contour lines corrrespond to the zero values of  the two fields.  The target field is Gaussian smoothed with a kernel of c $5\Mpch$.}
    \label{fig:target}
\end{figure}

\cref{fig:target} presents the ``target'' density and velocity field, from which the mock data has been drawn and which the \Hamlet\ algorithm is designed to recover. The linear over-desnity ($\delta$) and the radial component of the velocity field ($v_{\rm r}$) are depicted. The target field is smoothed with a Gaussian kernel of a radius of $5\Mpch$.

\begin{figure}
    \centering
    \includegraphics[width=\columnwidth]{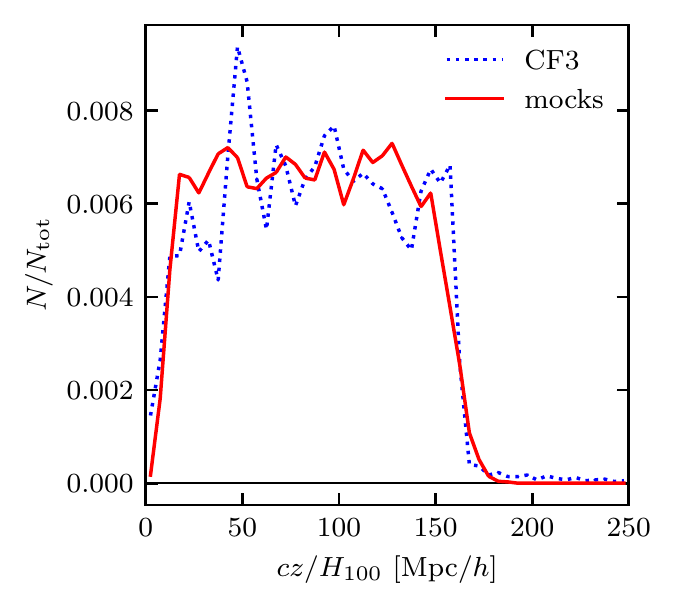}
    \caption{The ``selection function'' i.e. the distribution of the distances for the sample of points used as constraints. For comparison we also show CF3.}
    \label{fig:dist_cz}
\end{figure}

The constraints are   isotropically  selected within a sphere of radius of $160\Mpch$. A radial selection function is imposed so as to have a uniform distribution per radial distance bins, $P(d) = {\rm constant}$, where distances are measured with respect to the mock observer at the relative centre of the box. This choice is
motivated by the relative flatness of the redshift distance distribution of the CF3 grouped data points ( \cref{fig:dist_cz}). The $160\,\Mpch$   cut corresponds to the effective distance cut of the CF3  data. The errors assigned to the mock data points follow the redshift distribution of the errors of the actual CF3 data. The following procedure is used. Given a mock point it inherits the error of the actual CF3 data point that is closest to it in redshift.

The two main factors that affects the quality of the Bayesian reconstruction in general and the HMC in particular are the numbers of the data points and their associated errors. In the limit of very dense sampling of the data points and negligible errors the target field should be  reconstructed  with high fidelity. In the other  extreme case of very sparse sampling and large observational uncertainties, the null field predicted by the prior PDF is recovered. An ensemble of 9 different mock databases has been constructed so as to investigate how these two factors affect the outcome of the \Hamlet\ reconstruction. Three different numbers of data points are selected, $(7.5,\ 15,\ 30)\times 10^4$. The distance moduli errors are gauged by an $\alpha$ parameter,  $\sigma_\mu=\alpha\sigma^{\rm CF3}_\mu$, where  $\sigma^{\rm CF3}_\mu$ is the  value for the actual CF3 survey error\footnote{Varying $\sigma_\mu$ is meant to mimic the different precision of different standard
candles since e.g. the error on a distance obtained from the Tip of the Red Giant Branch method is around 5\% while from scaling relations are closer to 20 \%. Thus a catalogue of just TRGB distances would correspond to $\sigma_\mu=0.25$ with respect to CF3 which is scaling relation dominated.}. In otherwords $\alpha=1$ corresponds to the typical errors associated with the inherent uncertainty in scaling relations (eg. Tully-Fisher) distance measures, while $\alpha=1/10$ is meant to mimic a catalogue constructed entirely with more accurate distance measures, like TRGB or SuperNovae.
The 9 mock databases are assigned the 3 different numbers of data points and 3 different $\alpha$ values (\cref{table:mocks}). The case of $1.5\times 10^4$ data points and $\alpha=1$ corresponds most closely to  the grouped CF3 data. 
                
\begin{table}
    \centering
    \caption{Mock catalogues and their characteristics}
    \label{table:mocks}
    \begin{tabular}{l|c|c}
        Name                         & N of points & $\alpha$, Factor on $\sigma_\mu$ \\
        \hline
        \hline
        CF3+ like                    & 15~000 &  1 \\
        Better measurements          & 15~000 & 1/2 \\
        Very good measurements       & 15~000 & 1/10 \\
        \hline
        More measurements            & 30~000 & 1 \\
        More, better measurements    & 30~000 & 1/2 \\
        More, very good measurements & 30~000 & 1/10 \\
        \hline
        Fewer measurements            & 7~500 & 1 \\
        Fewer, better measurements    & 7~500 & 1/2 \\
        Fewer, very good measurements & 7~500 & 1/10
    \end{tabular}
\end{table}

A note of caution on the expected effect of the  sharp drop has on   the Bayesian reconstruction is due here.
\cite{Hinton2017} have investigated this exact problem: the effect that a sample selection function with a sharp cutoff has on the likelihood function and thereby on the Bayesian posterior PDF. Their
conclusion is that the inferred variables close to the edge of the data, namely close to the  cutoff,  are biased. Our analysis and findings support the finding of  \cite{Hinton2017}. Consequently we limit our analysis and present results only within a sphere of a radius of $150\Mpch$.

\subsection{Convergence}
\label{sec:conv_bias}

A critical issue that MCMC methods in general and the HMC in particular  face is that of convergence, namely how long should an MCMC chain or HMC trajectory be   to meet some given criteria  of confidence in the estimated parameters. (The discussion that follows focuses on HMC trajectories but it implies to MCMC chains in general.) The HMC trajectories never 'rest' and keep on 'moving' in the parameters space. 
Three obvious issues to consider are: a. Does the HMC trajectory oscillate around the 'true' values of the parameters, namely the issue of bias; b. What is the scatter exhibited by the trajectory, i.e. the variance around the mean (defined by \cref{eq:int_mc}); c. What is the rate of convergence. The convergence rate is discussed here and the issues of bias and variance are addressed in subsections \ref{sec:reconstruction} and \ref{sec:correlation} below.

\begin{figure}
    \centering
    \includegraphics[width=\columnwidth]{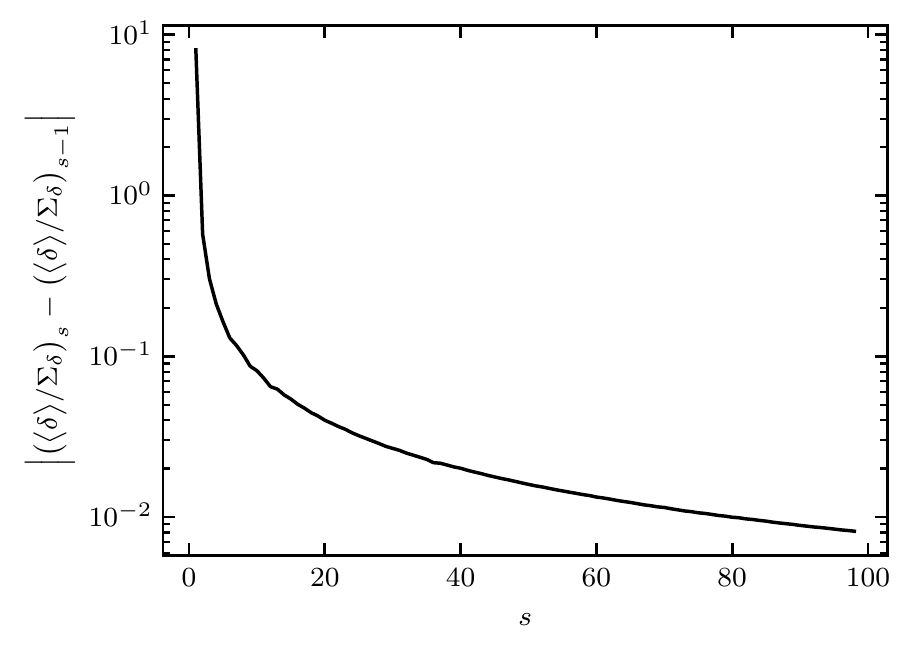}
    \caption{Differences between successive partial estimations of the signal over noise ratio field
    as a function of the number of steps used to estimate it. All mocks converge in a very identical
    way. }
    \label{fig:conv_dsnr}
\end{figure}

The rate of convergence is examined here by monitoring the change of the mean of the density field, $\big<\delta\big>$, along the HMC trajectory. \cref{fig:conv_dsnr} presents the differential change in  $\left<\delta\right>$ normalized by the  (square root of the) cosmic variance between successive steps, $\left|  \left(\left<\delta\right>/\Sigma_{\delta}\right)_s  - \left(\left<\delta\right>/\Sigma_{\delta}\right)_{s-1} \right| $. Here $\Sigma^2_{\delta}$ is the variance of the $\delta$ field evaluated at a given step of the HMC. \cref{fig:conv_dsnr} indicates that to get to percent level convergence requires on the order of 100 chains.

\subsection{Reconstruction of the large scale structure}
\label{sec:reconstruction}

The \Hamlet\ method's main mission is to perform a Bayesian estimation of the LSS, namely the density and velocity fields, from Cosmicflows-like databases. To meet that end and examine it we focus here on a subset of the $\vq$ parameters, 
$\vq_{\rm LSS} = \left(\delta_{k,1}^R,\, \hdots ,\,\delta_{k,m}^R,\, \delta_{k,1}^I,\, \hdots ,\,\delta_{k,m}^I \right)$, 
namely the ones that determine the LSS. The conditional mean field, given the data and the prior model is readily written (as a particular case of the general \cref {eq:int_mc}) as:
\begin{equation}
\vq^{\rm mean}_{\rm LSS} =	\mean{\vq_{\rm LSS} \ | {\rm\ data,\ prior}}  \approx \frac{1}{n_S} \sum_{s=1}^{n_S} \vq_{\rm LSS,s}
    \label{eq:q_LSS_mean}
\end{equation}
An ensemble of $n_{\rm IS}$ independent states (IS) of the HMC chain is  constructed as well $\left\{ \vq^{\rm IS}_{{\rm LSS},i} \right\}\ \left(i=1, \hdots , n_{\rm IS}\right) $. The conditional mean field and the  ensemble of ISs are the HMC equivalent of the WF estimator  and the ensemble of CRs of the WF/CRs algorithm.

\begin{figure*}
    \centering
    \includegraphics[width=\textwidth]{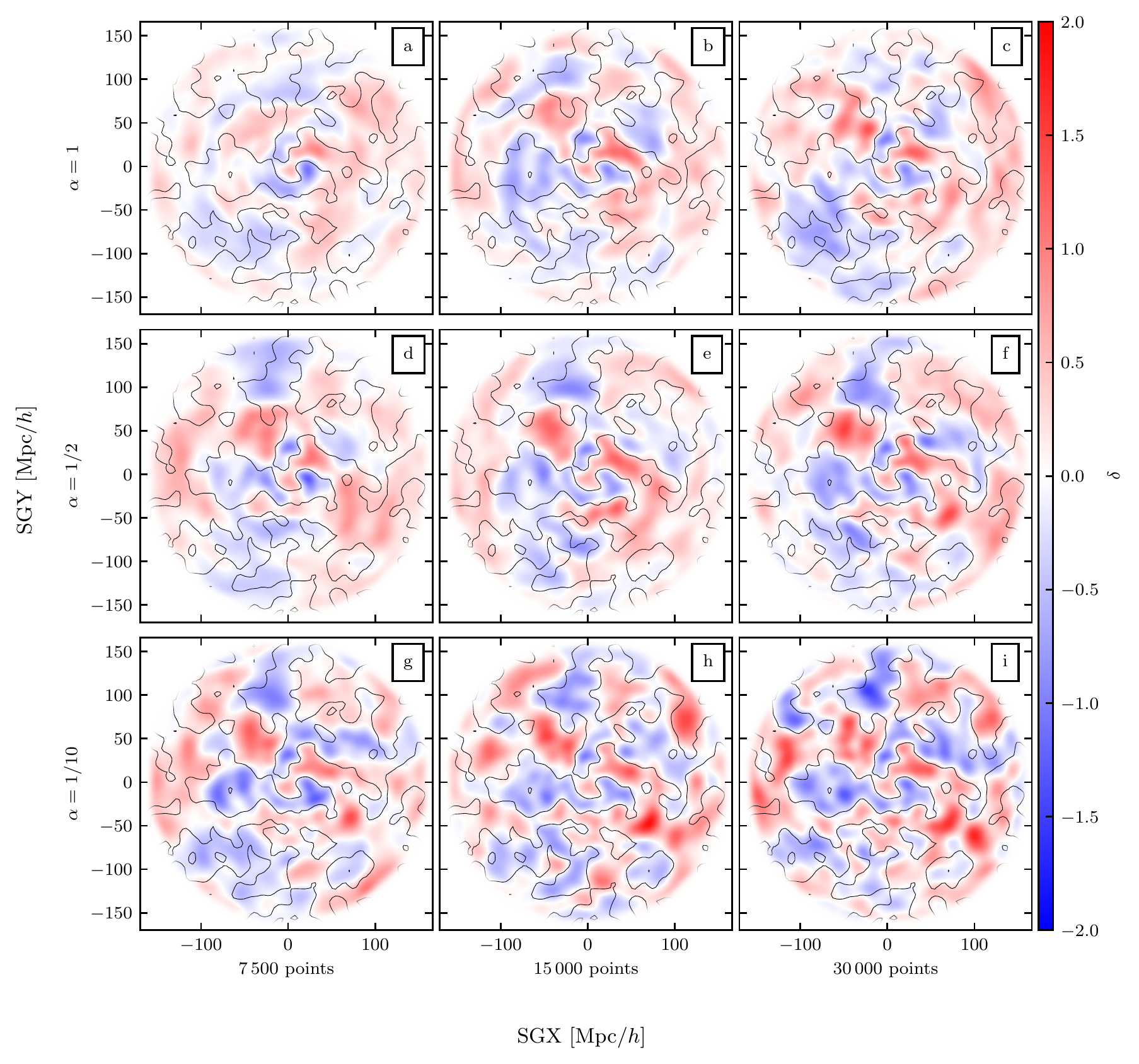}
    \caption{The conditional mean $\delta$ field given CF3-like mock data and the \LCDM\ prior model. The field is Gaussian smoothed with a kernel of $5\Mpch$. The plots show color maps of a slice of the $\delta$ field. The color bar indicates the color coding of $\delta$ and the contour lines correspond to the zero level. The frames correspond to the different mock catalogs dented by $(N/10^4, \alpha$, where $N$ is the number  of data points and $\alpha$ controls the level of the observational errors, $\sigma_\mu=\alpha\sigma^{\rm CF3}_\mu$. The nine mock data are: 
    a. (0.75, 1.0), b. (1.5, 1.0), c. (3., 1.0); 
    d. (0.75, 0.5), e. (1.5, 0.5), f. (3., 0.5); 
    g. (0.75, 0.1), h. (1.5, 0.1), i. (3., 0.1).
    Frame {\bf b} corresponds to the actual CF3 data in terms of the number of data points and the magnitude of the errors.}
    \label{fig:mean_divv}
\end{figure*}

\cref{fig:mean_divv} shows a slice of the $\delta$ field for the 9 different mock data sets (see \cref{table:mocks}). The presented density field is the conditional mean field given the data, namely   $\delta^{\rm mean}=\delta\left(\vq^{\rm mean}_{\rm LSS} \right)$.  The grid of density maps reflects the change of quality of the data - more data points and smaller errors corresponding to better data. Degradation of the data leads to an attenuation of  $\delta^{\rm mean}$. This a manifestation of the well known property of the Bayesian estimation - the worse the data the  more biased the results are towards the null field predicted by the prior model \citep[cf.][]{1995ApJ...449..446Z}. The grey lines represent the $\delta=0$ contour of the target. The reader will note how similar these are to the the limit of small errors and large data set (i.e. \cref{fig:mean_divv}(i)) Indeed, in general, the  target and reconstructions $\delta=0$ contours match for $\alpha=1/10$, \cref{fig:mean_divv}(g,h,i). In the case of the CF3 mock ($\alpha=1$, 15,000 points, figure  \cref{fig:mean_divv}(b) ), the target density contour is fairly accurately recovered in the inner regions, ie within around $50-60\,\Mpch$. Examining the density field reconstructions ``vertically'' indicates that the single most important factor for obtaining an accurate density field reconstruction is the data quality. Namely: {\it Better data is more important than larger data sets}. For a given catalogue size, better data allows the reconstruction to be accurate at greater distances. For a given error, more data improves the reconstructions at fixed distances, instead of extending the improvement of the reconstruction.

\begin{figure*}
    \centering
    \includegraphics[width=\textwidth]{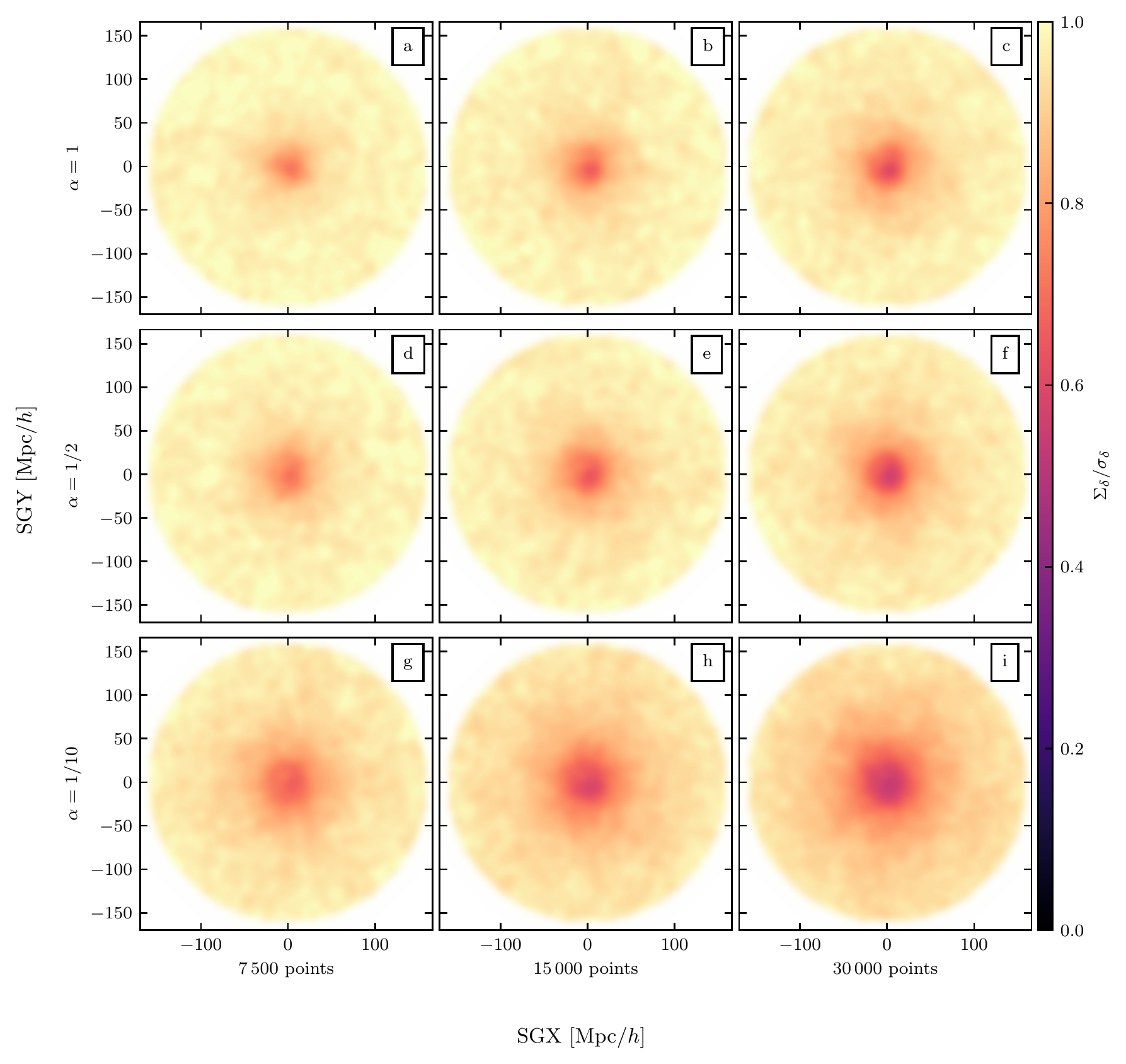}
    \caption{Color maps of the local constrained variance normalized by the cosmic variance of the $\delta$ field,  $\sigma_\delta(\vr)/\Sigma_\delta$. The conventions and structure of \cref{fig:mean_divv} are followed here, 
    }
    \label{fig:std_divv}
\end{figure*}

The linear over-density field constitutes a random Gaussian field whose variance is determined by the power spectrum of the field and the resolution of the given realizations of the field. This is the cosmic variance of the $\delta$ field, denoted by $\sigma_\delta^2$. The variance of different states along the HMC chain, $\Sigma_\delta^2$,  varies according to the `strength` of the data and the properties of prior model. Furthermore, it varies with the location at which it is evaluate. i.e. $\Sigma_\delta=\Sigma_\delta(\vr)$. The spatial variation of $\Sigma_\delta(\vr)$ reveals the constraining power of the data, given the prior model. Where $\sigma_\delta(\vr)/\Sigma_\delta \ll 1$ the density field is strongly constrained by the data and the prior model and only very small scatter is expected to be found around the mean field. Where $\Sigma_\delta(\vr)/\sigma_\delta \sim 1$ the field id essentially unconstrained by the data and the prior model. \cref{fig:std_divv} presents the variation of normalized constrained variance, $\Sigma_\delta(\vr)/\sigma_\delta$, for the 9 mock databases presented in \cref{fig:mean_divv}. Inspection of \cref{fig:std_divv} reveals that for all the mock data considered the quality of the reconstruction degrades with the distances from the (mock) observer. This is a reflection of the degradation of the data with the distance - the magnitude of the errors increases and the density of data point decreases with distance. The picture of how the reconstruction degrades with distance as a function of date set size and error reinforces the conclusions drawn from \cref{fig:mean_divv} namely smaller errors on the data improves the reconstructions more than larger data sets. Small data sets with small errors are worth more than large data sets with large errors when examining the reconstructed density field.

\begin{figure*}
    \centering
    \includegraphics[width=\textwidth]{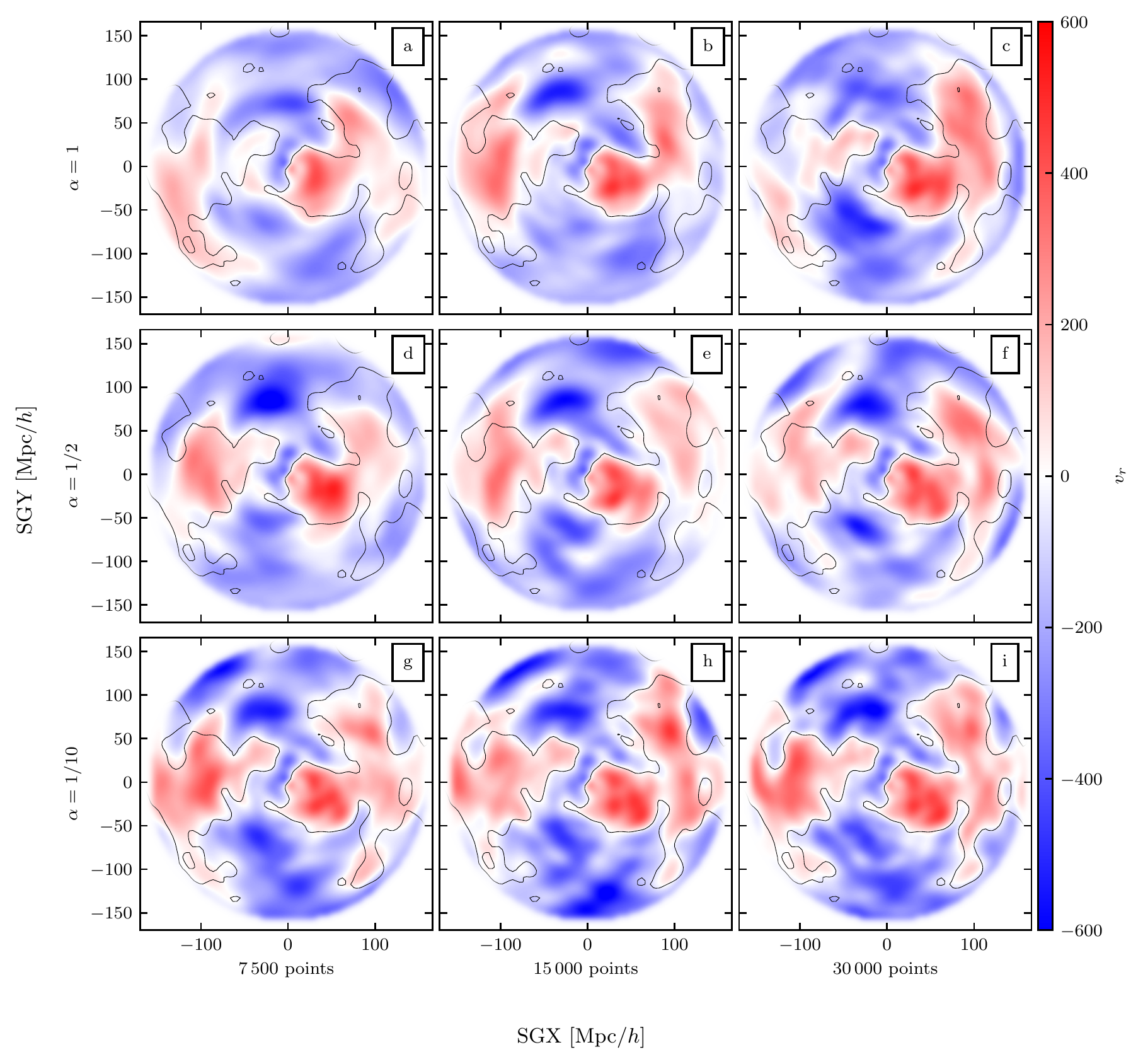}
    \caption{Same as \cref{fig:mean_divv} but for the radial component of the  velocity field.}
    \label{fig:mean_vr}
\end{figure*}

\begin{figure*}
    \centering
    \includegraphics[width=\textwidth]{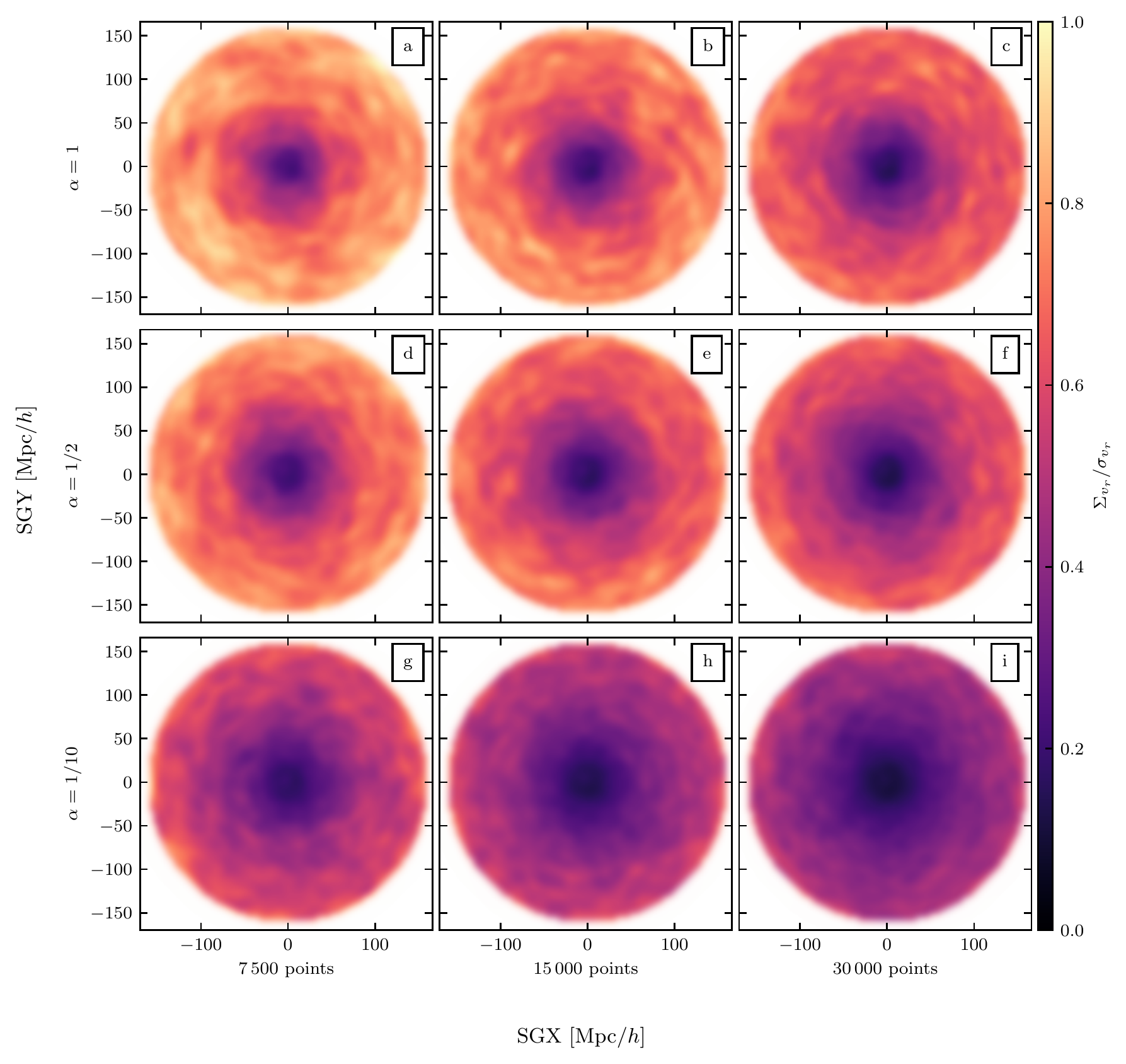}
    \caption{Same as \cref{fig:std_divv} but for the radial component of velocity field ($v_{\rm r}(\vr)$). The constrained and comic variances are evaluated for $v_{\rm r}(\vr)$.  }
    \label{fig:std_vr}
\end{figure*}

Next, the radial component of the velocity field is investigated (\cref{fig:mean_vr,fig:std_vr}). The velocity field power spectrum is `redder' than that of the $\delta$ field, namely it has more power on long wavelength compared with the short one, hence the velocities effective correlation length is larger than that of the densities. Hence one expects the velocities to be more constrained by the data than the densities. This is clearly manifested by  \cref{fig:mean_vr}. A visual of how well the target's $v_{r} = 0$ contour matches the reconstructed radial velocity field indicates that even in the case of CF3 like mock (ie  \cref{fig:mean_vr}b) The reconstructed velocity field is doing a good job at greater distances (e.g as compared with the $\delta$ field). The velocity field around large, distance concentrations of matter and voids is accurately reconstructed with \Hamlet. We can qualitatively asses the superiority of the velocity field reconstruction as compared to the density field by examining  \cref{fig:std_vr} (and comparing to  \cref{fig:std_divv}) which shows just how well similar the reconstructed velocity field is to the target.

\subsection{Monopole and dipole}

Global measures of the velocity field are given  by the volume-weighted mean monopole and dipole moments of the velocity field in spheres of radius $R$ \citep[for details]{2021MNRAS.505.3380H}.
The monopole moment is  the  mean of  $-\nabla\cdot\vv/H_0$, where the  scaling by $H_0$ is introduced so as to make the expression dimensionless and proportional to the mean (linear) over-density within $R$. The minus  sign is introduced so as to make   the monopole within a sphere of $R$  to be proportional to the mean  over-density within  that volume. The dipole moment is the (volume weighted) mean value of the velocity, namely it is the bulk velocity of a sphere of radius $R$. The variation of the monopole and dipole moments with depth provides a global measure of the underlying LSS of the universe, and as such they serve as good monitors of the quality of the \Hamlet\ reconstruction.

\begin{figure*}
    \centering
    \includegraphics[width=\textwidth]{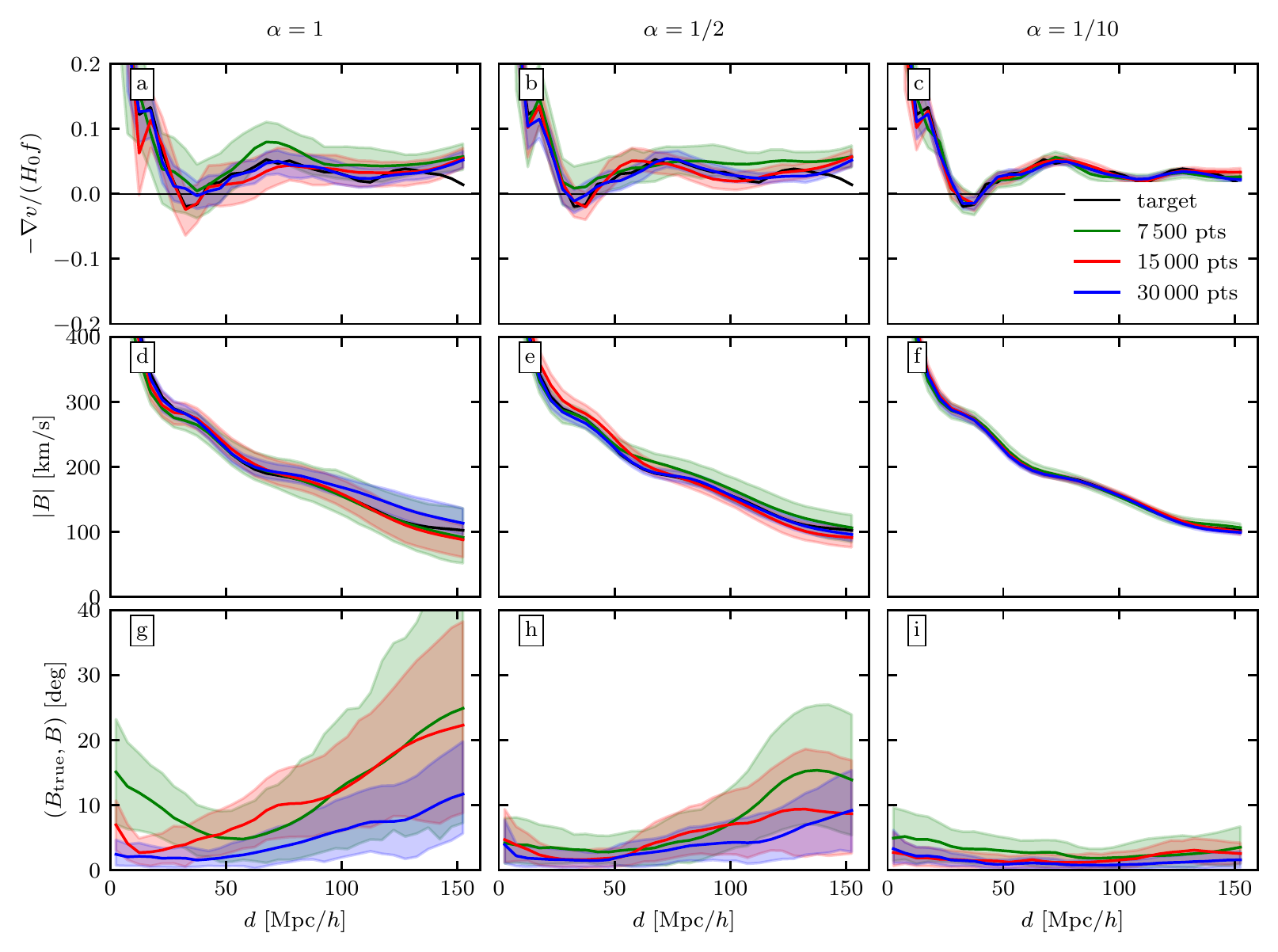}
    \caption{Top row: The monopole moment (namely the mean fractional over-density; upper panel), the amplitude (middle panel) and alignment (with the target; lower panel) of the dipole moment (namely the bulk velocity)   are shown  shown for $\alpha =1, 1/2$ and $1/10$ (columns left to right) and for $N = \big(0.75$ (green),  $1.5$ (red)  and $3.0$ (blue)$\big)\times10^4$, where $N$ is the number of data points. The mean (solid lines) and the scatter are calculated over the ensemble of HMC steps.  The moments of the target field are shown in black. }
  \label{fig:flows}
\end{figure*}

\cref{fig:flows} presents the variation with depth of the monopole (upper row) and the dipole (middle row) of the velocity field, namely the mean fractional over-density  ($\delta$) and the bulk velocity of a sphere of radius $R$. The lower row shows the alignment of the bulk velocity of the reconstructed and the target velocity fields. The dependence of the estimation of the radial profiles of the moments on the quality of the data is investigated. The profiles are shown as a function of the magnitude of the errors ($\alpha=\left(1.0,\ 0.5,\ {\rm and} \ 0.1\right)$ and of 
the number of data points, $N = \left(0.75,\ 1.5\ {\rm and}\ 3.0 \right)\times10^4$. The profiles are presented by their mean and standard deviation taken over the ensemble of steps.

The plots of  \cref{fig:flows} are informative. With the exception of the behaviour of the monopole moment close to the edge of the data, $R\lessapprox150\ \Mpch$, where the reconstructed monopole exceeds that of the target one. The edge-of-the-data discrepancy is in line with the findings of \cite{Hinton2017}. As for all the other cases they behave as expected. The mean \Hamlet\ profiles deviation from the target profile and the scatter around the mean profiles  grow with $R$ and get smaller with the increase of the number of steps. A note is due here on the large scatter in the amplitude and alignment  of the bulk velocity at $R\gtrapprox 100\ \Mpch$, say. The bulk velocity of a sphere of radius $R$ is induced by structures outside that radius. The target field is constructed within a box of side length 0f $L=500\ \Mpch$ with periodic boundary condition, which renders the power within the box and outside these radii, and its constraining power to be rather small. This is manifested by the large scatter around the mean. The lesson to be learnt here is that the reconstruction of the bulk velocity on a given scale $R$ needs to be  done within boxes of $L$ much larger than $R$.

\subsection{Correlation Coefficients}
\label{sec:correlation}
\begin{figure*}
    \centering
    \includegraphics[width=\textwidth]{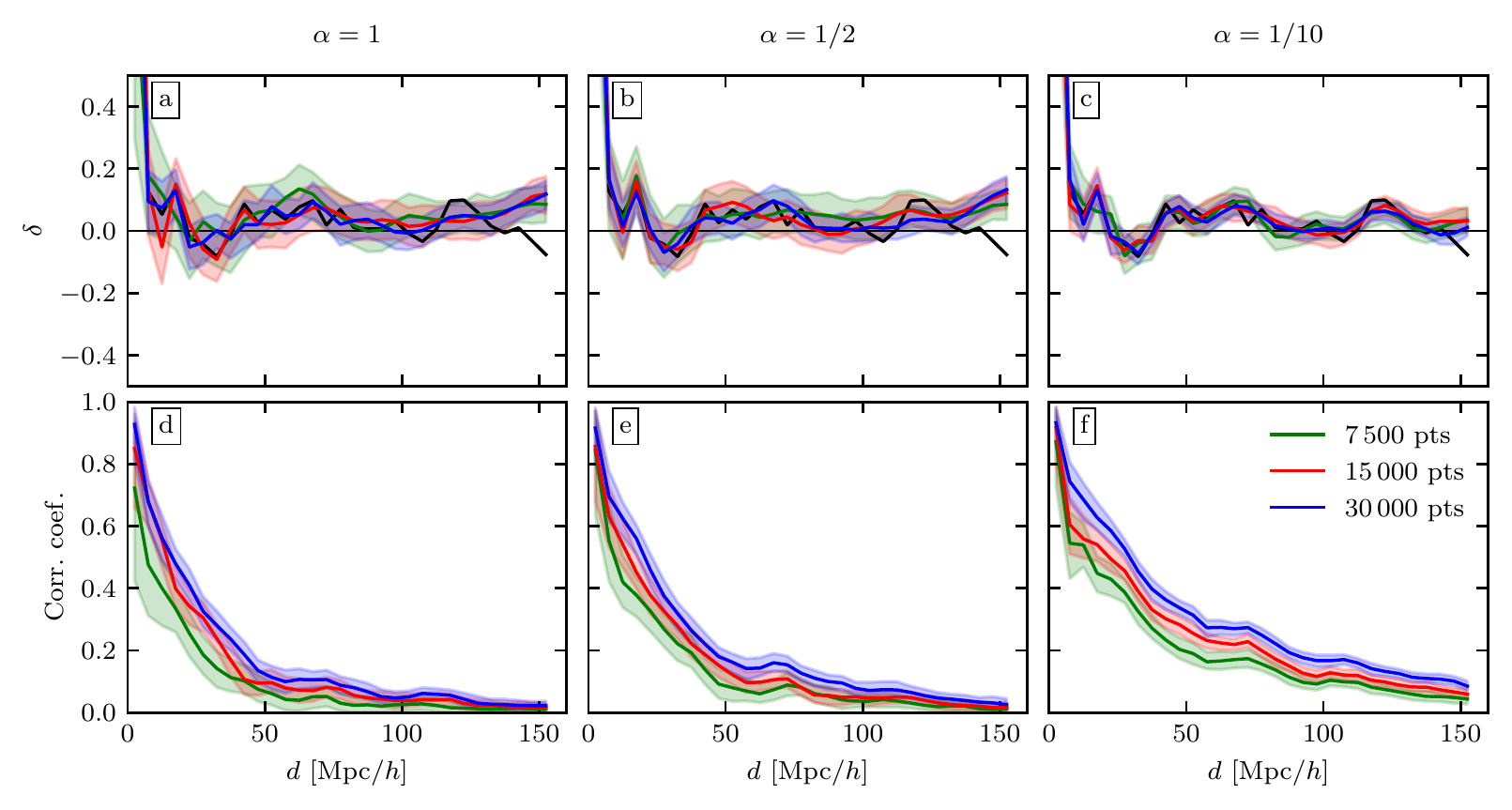}
    \caption{Top row: The mean plus-minus standard deviation of the density field per distance shell
        is shown for $\alpha =1, 1/2$ and $1/10$ (left to right). In black we show the target
        results while in green, red and blue we show the curves reconstructed from 7\,500, 15\,000
        and 30\,000 points respectively. Beyond this region there are no constraints. Bottom row:
        same plots but for the correlation between the reconstructed field and the target field.}
    \label{fig:mean_corr_divv}
\end{figure*}

 Next, the fidelity of the  \Hamlet\ reconstruction is monitored by means of the scatter  of the density and radial velocity fields evaluated in radial shells. The upper panel of \cref{fig:mean_corr_divv} shows the mean and scatter of the density in spherical shells of width of  $\Delta R=10\ \Mpch$ for the 3 values of $\alpha$ and the 3 assumed sizes of the mock catalogue. The mean density of the target field is presented for reference. The disagreement at $R\lessapprox150\, \Mpch$ is again clearly manifested. Next the density field within the shells is examined by studying the correlation between the densities of individual voxels (grid cells) within the shells. The lower panel of \cref{fig:mean_corr_divv} shows the (Pearson) correlation coefficients  of the \Hamlet\ reconstructed and the target densities. The correlation coefficient profiles are again plotted for the 3 levels of errors and the 3 ensembles of the HMC chain.  In general, the \Hamlet reconstruction is follows the target quite well. In the limit of much data and small errors, the reconstruction is very close to the target.                 
 
\begin{figure*}
    \centering
    \includegraphics[width=\textwidth]{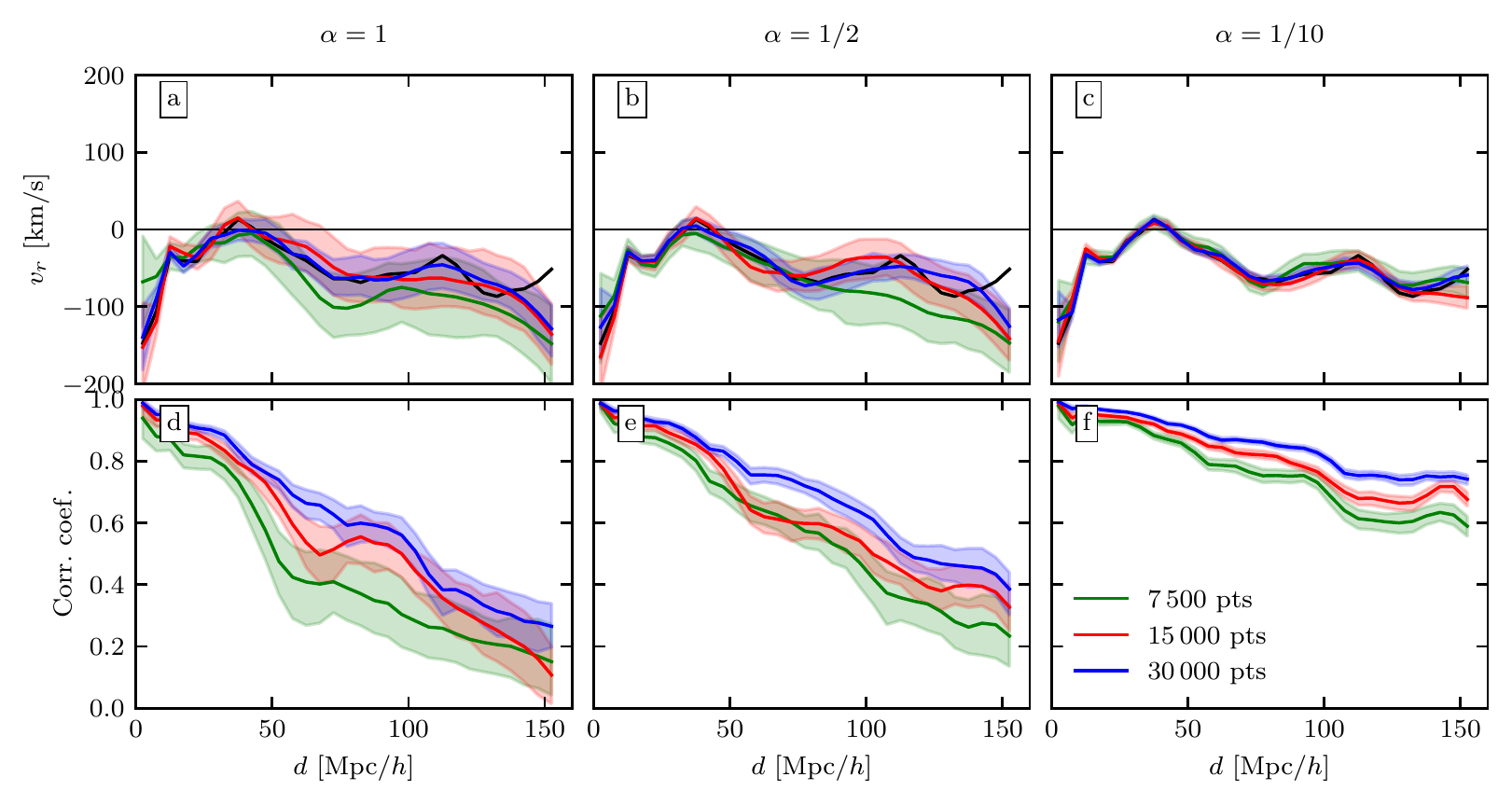}
    \caption{Same as \cref{fig:mean_corr_divv} but for the radial velocity field instead of the
    density field.}
    \label{fig:mean_corr_vr}
\end{figure*}

 \cref{fig:mean_corr_vr} applies  the analysis of  \cref{fig:mean_corr_divv} to the case of the radial velocity field. The mean and scatter of $v_r$ within spherical shells (upper row) and the correlation coefficient with the target field (lower row) are presented in \cref{fig:mean_corr_vr}. As expected, the comparison of the 2 figures clearly shows that the (radial) velocity field is much more correlated than the density field. Smoothing the target and the reconstructed fields would make them much more correlated. Again, in the limit of much data and small errors the reconstructed velocity field is so accurate that a correlation coefficient of nearly 0.8 is obtained at $150\,\Mpch$. Obtaining a CF like catalogue with such large numbers of data points is already a reality in CF4. On the other hand, having small errors corresponding to $\alpha =1/10$ is still slightly unrealistic, but not unimaginable with future purpose built telescopes designed specifically to monitor variable stars at cosmological distances.

\section{Summary and discussion}
\label{sec:summary}

The problem of the reconstruction of the large scale density and velocity fields from peculiar velocities surveys is addressed here within a Bayesian framework. In particular, the reconstruction aims at  Cosmicflows-like data where observational uncertainties are on the distance moduli, which results in a lognormal bias on the estimated distances and velocities.   The HAmiltonian Monte carlo reconstruction of the Local EnvironmenT (\Hamlet\ algorithm performs the reconstruction within the framework of the linear theory of the \LCDM\ standard cosmological model, which is taken here as the Bayesian prior, using the Hamiltonian Monte Carlo (HMC) method to sample the posterior probability distribution function (PDF) given the \LCDM\ model and the Cosmicflows-like data. Like previous MCMC treatments of the problem \citep{2016MNRAS.457..172L,Graziani2019} the \Hamlet\  samples the posterior PDF of true distance of the data points coupled with the underlying linear density field. This   differs from the Wiener filters and constrained realizations (WF/CRs) approach where the correction of the lognormal bias is done independently of the Bayesian reconstruction of the LSS \citep{2015MNRAS.450.2644S,2021MNRAS.505.3380H}.

The current \Hamlet\ HMC algorithm and the \cite{2016MNRAS.457..172L,Graziani2019} MCMC ones are formulated within the same mathematical Bayesian framework, making similar assumptions on the prior PDF and deriving the same posterior PDF from the same input data. The main difference between the standard (e.g. Metropolis-Hastings or Gibbs sampling) MCMC algorithm and the HMC is in the sampling of the posterior PDF. The extremely low rejection rate of the HMC steps and its ability to be run on GPUs, makes the procedure very efficient. A comparison of the performance of the \Hamlet\ algorithm with the one presented in \citet{Graziani2019} finds an efficiency gain factor of two to four orders of magnitude in favor of the HMC. This gain in efficiency will enable a very significant increase of resolution in future applications of the \Hamlet\ algorithm compared to to the resolutions used in the MCMC cases  \citep{2016MNRAS.457..172L,Graziani2019}.

The successful application of the \Hamlet\ algorithm to Cosmicflows-like mock data  paves the way for its application to the actual Cosmicflows-3 data \citep{2016AJ....152...50T} and the upcoming next data release of Cosmicflows. Four immediate specific goals that can be achieved  by such an applications are: 
1. The reconstruction of the present epoch large scale velocity field, including the bulk velocity of the local volume; 
2. The construction of constrained initial conditions of cosmological simulations, following earlier studies that have the WF/CRs algorithm \citep[e.g.][]{2014NewAR..58....1Y,2016MNRAS.455.2078S,Libeskind2020}; 
3. Adding to the list of parameters of the Bayesian model (\cref{eq:q})  cosmological parameters such as $H_0$ and the $\sigma_8$ normalization of the power spectrum, and estimating these self-consistently from data;
4. Studying the internal consistency of the different components of the Cosmicflows data, such as the zero-point calibration of the different data types of the Cosmicflows. This can be done by assigning different Hubble parameters to the different data sub-sets, and estimating them by Bayesian inference.

A few days before the submission of the current paper we came across the  
\cite{2021arXiv211115535B} preprint on the arXiv. The current  and  the \citeauthor{2021arXiv211115535B}'s  studies are similar - both address the problem of the reconstruction of the LSS from peculiar velocities surveys by means of HMC algorithms. The studies have been carried out independently and have been applied to different data sets. A detailed comparison of the two will be presented elsewhere.

\section*{Acknowledgements}
Useful discussions with Tamara Davis, concerning the \cite{Hinton2017} paper, are acknowledged.
This work has been done within the framework of the Constrained Local UniversE Simulations (CLUES) simulations. AV and NIL acknowledge financial support from the Project IDEXLYON at the University of Lyon under the Investments for the Future Program (ANR-16-IDEX-0005).
YH has been partially supported by the Israel Science Foundation grant ISF 1358/18.

\section*{Data availability}
 	
The data underlying this article will be shared on reasonable request to the corresponding author. 

\bibliographystyle{mnras}
\bibliography{HMC}

\end{document}